\documentclass[iop,apjl]{emulateapj} %  2 colunas

\usepackage[usenames]{color}

\usepackage{soul}
\usepackage{color}

\slugcomment{Submitted to ApJL. Manuscript  LET00000 }

\newcommand{\blue}{\textcolor{blue}}

\newcommand{\teff}{$T_{\rm eff}$}

\newcommand{\msun}{$\mathrm{{\it M}_{\odot}}$}

\def\ltsim{\raise 1.0pt \hbox {$<$} \kern-0.8em \lower 2.0pt \hbox {$\sim$}}

\shorttitle{Solar twin candidates revealed by the \emph{Kepler}  mission}

\shortauthors{do Nascimento et al.}

\begin{document}

\title{Rotation periods and ages of solar analogs and solar twins revealed by the \emph{Kepler} mission}

\author{J.-D.~do Nascimento, Jr.$^{1,2}$, R. A. Garc\'ia$^3$,  S. Mathur$^4$, F. Anthony$^2$, S. A. Barnes$^{5}$, S. Meibom$^{1}$, J.S.~da Costa$^2$, M. Castro$^2$, D.  Salabert$^3$ and T.  Ceillier$^3$}

\altaffiltext{1}{Harvard-Smithsonian Center for Astrophysics, Cambridge, MA 02138, USA; jdonascimento@cfa.harvard.edu}

\altaffiltext{2}{Universidade Federal do Rio Grande do Norte, UFRN, Dep. de F\'{\i}sica Te\'orica e Experimental,
DFTE, CP 1641, 59072-970, Natal, RN, Brazil}

\altaffiltext{3}{Laboratoire AIM, CEA/DSM -- CNRS -, Univ. Paris Diderot -- IRFU/SAp, Centre de Saclay, F-91191 Gif-sur-Yvette Cedex, France}

\altaffiltext{4}{Space Science Institute, 4750 Walnut Street Suite 205, Boulder CO 80301, USA}

\altaffiltext{5}{Leibniz-Institute for Astrophysics, Potsdam D-14467, Germany}

%\slugcomment{Send proofs to:  J.D. do Nascimento}

\begin{abstract}
{A new sample of  solar analogs and twin candidates have been constructed and studied, with particular
attention to their light curves from NASA's \emph{Kepler} mission.  This letter aims to assess the 
evolutionary status,  derive their rotation and ages and  identify  those solar analogs or 
solar twin candidates. We separate out the subgiants that compose
a large fraction of the asteroseismic sample, and which show an increase in the average rotation period
as the stars ascend the subgiant branch. The rotation periods of the dwarfs, ranging  from 6 to 30 days,
and averaged 19d, allow us to assess their individual evolutionary states on the main
sequence, and to derive their ages using gyrochronology. These ages are found to be in agreement
with a correlation coefficient of r = 0.79  with the independent asteroseismic ages, where available.
As a result of this investigation, we are able to identify 34 stars as solar analogs 
and 22 of them as solar twin candidates.}

  \end{abstract}


%________________________________________________________________

\section{Introduction}
\label{intro}
The Sun is a benchmark in stellar astrophysics research and establishing a sample of  solar analog stars 
is important to map its past, present and future.  About 50 yr ago, Olin Wilson and collaborators discovered 
Sun-like activity  cycles in a group of 
$\sim100$ stars,  now known as the Mt.\,Wilson sample (\blue{\citealt{wilson1963}}). Additional work 
over the intervening  decades    (e.g., \blue{\citealt{Noyes84}}) has made us confident  that  some of  
these stars do indeed have ages comparable to that  of the Sun and that their activity, and chemical and  
other fundamental properties make them, in  many ways,  ``solar analogs''.  Many additional studies  
have allowed the identification of some other stars as solar analogs and some of these  have
properties that are close enough to those of the Sun that they are even called ``solar twins''.   
Solar twins  are  spectroscopically  indistinguishable from the Sun \blue{\citep{cayrel1996}}
for example,  18~Sco (\blue{\citealt{portomello1997}}; \blue{\citealt{bazot}}),  
CoRoT Sol 1 \blue{\citep{donascimento2013}}, and HIP~102152 (\blue{\citealt{Monroe13}}). 
 The term ``solar analogs" here refers to stars with  
0.9 $<$ $M$/\msun~$\leq$~1.1  and ``solar twin candidates"  refers to stars with  
0.95 $<$ $M$/\msun~$\leq$~1.05 and   rotation periods of $P_{\mathrm {rot}}$ $>$ 14 days.

 \vspace{0.2cm}

Solar twins also allow us to decide to what extent the Sun itself can be considered  a 
``typical''  1.0~\msun~ star  \blue{{\citep{gusta2008}}}. The search for solar twins has been 
greatly  expanded since 1997 \blue{\citep{hardorp1978, cayrel1996}}, when only one solar
 twin was  known, and currently  more than two dozen  twins have been  identified    
 \blue{\citep[e.g.,][]{melendez2007,donascimento2013,Monroe13}}.
Solar twins are also important to calibrate fundamental UBV(RI)$_{C}$ --- \teff~relations  
\blue{{\citep{portomello1997, ivan2012}}} and to test non-standard stellar  models \blue{\citep[e.g.,][]
{bazot}}. A sample of solar twins with determined $P_{\mathrm {rot}}$ is  also important to study the
``Sun in Time''  (see  \blue{\citealt{dorren_guinan1994}}). 

 \vspace{0.2cm}

 A related consideration is  how the Sun has changed over time, and how its
behavior relates to that of other younger and  older stars of solar mass. Studying this requires 
that we assemble a sample of stars whose ages we know  well enough to place them all in a time sequence
that includes the Sun.  However, ages for field stars, particularly for those on the main sequence,  are notoriously difficult to
derive (e.g., \blue{\citealt{Barnes07}}; \blue{\citealt{Soderblom10}}).  Consequently, the classical distinction between 
solar-type and analogs or twins does not include age constraints apart from the obvious dwarf/giant distinction. 
\begin{figure*}   
\centering
\vspace{0cm}
\hspace{-1.2cm}
\includegraphics[angle=0,width=8.5cm,height=8.5cm]{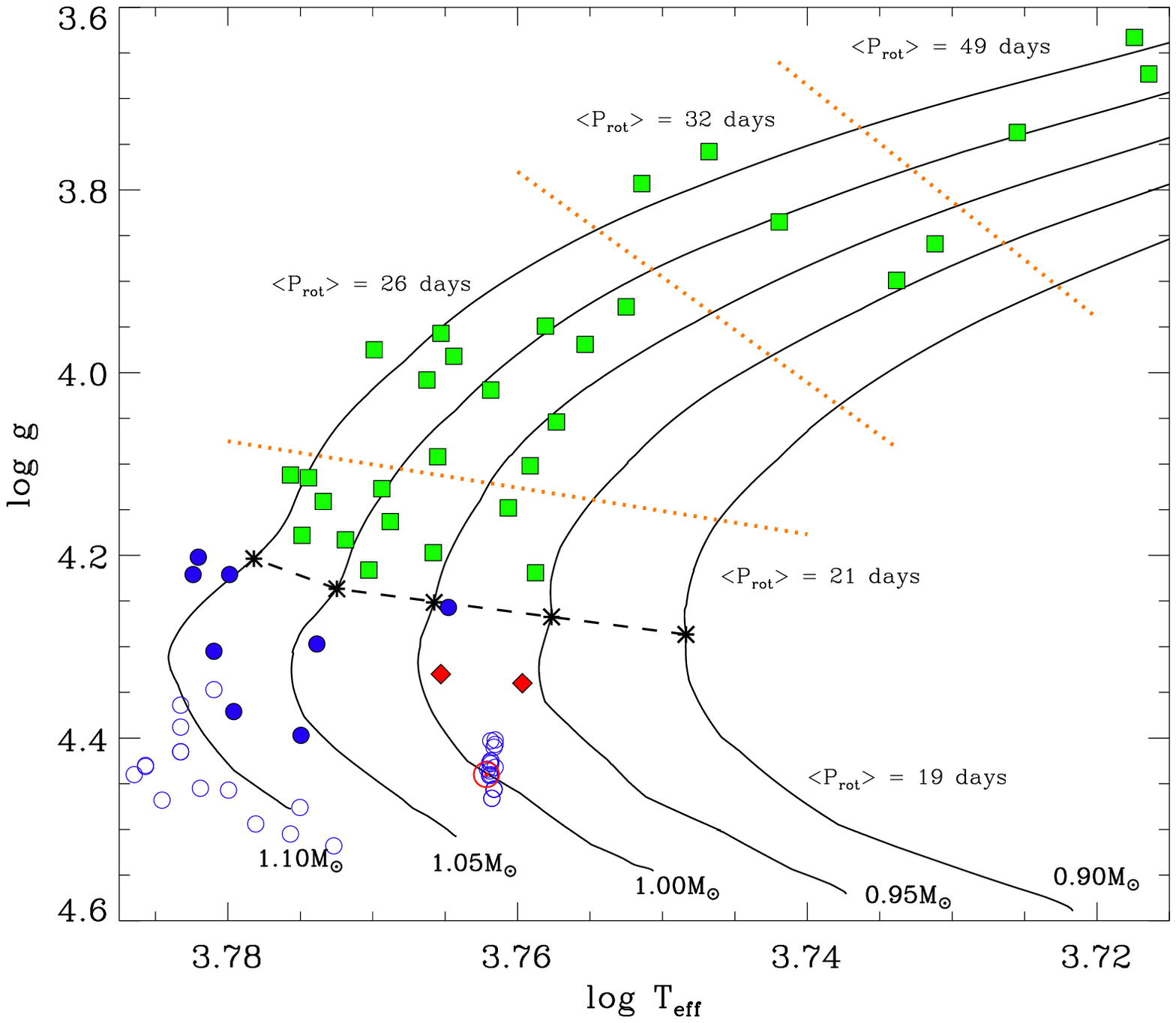}
\includegraphics[angle=0,width=8.5cm,height=8.5cm]{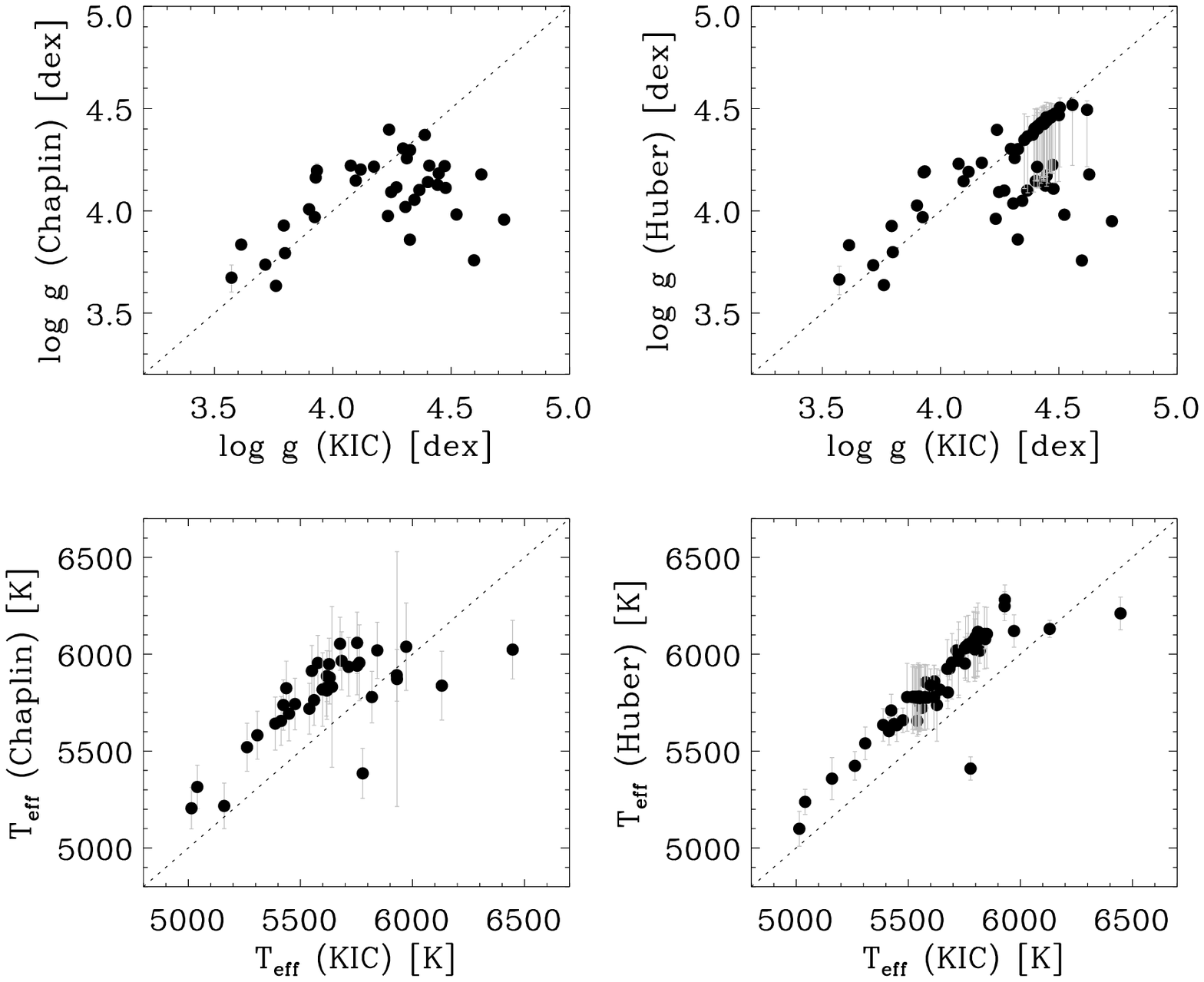}
\caption{Large panel on the left shows the $\log g$ - \teff~ for the entire  
sample.  Open circles  represent dwarfs selected from the KIC  and filled circles dwarfs from the seismic sample.
Squares represent subgiants.  Diamonds  indicate 16~Cyg~A and B 
evolutionary status. The upper panels on the right compare the $\log g$ values of
\blue{\cite{chaplin2014}} and \blue{\cite{huber2013}}  to the KIC values. The bottom
panels of the right compare the corresponding \teff~values. }
\label{evoldig_f1}
\vspace{0.2cm}
\end{figure*}
The various requirements make finding solar analogs and twins difficult.  A good way to find such valuable 
stars is to take advantage of space missions such as \emph{CoRoT} \blue{{\citep{baglin06}}}  and NASA's \emph{Kepler} mission
 \blue{{\citep{Borucki}}}. These missions have provided precise  observations for thousands of 
 main-sequence stars. Furthermore, we can detect periodic stellar variability for these stars. 
 Such modulation is a signature of the presence of spots on the star's
surface and can be used to measure the stellar $P_{\mathrm {rot}}$  \blue{\citep[e.g.,][]{bas+11, mei+11,  
donascimento2013, McQuillan2013}}. From $P_{\mathrm {rot}}$,  ages can be estimated 
using gyrochronology  \blue{\citep{Barnes07}}. 

 \vspace{0.2cm}

Although recent works have increased the number of solar twins 
and studied their physical parameters and chemical abundances in detail, 
their $P_{\mathrm {rot}}$ are mostly unknown, except for the two solar twins 
18 Sco \blue{\citep{portomello1997}} and CoRoT Sol 1 \blue{\citep{donascimento2013}}.
\section{Working sample}  % Sect. 2 %%%%%%%%%%%%%%%%%%%%%%
\label{sel}
Our  sample  of 75 stars consists of a  {\it seismic sample} of 38 from  
\blue{\citet{chaplin2014}}, 35 additional stars  selected from the \emph{Kepler} Input Catalog (KIC), and 16~Cyg~A and  B. 
We selected 38 well-studied stars from the asteroseismic data with  fundamental properties, including ages,  estimated by \blue{\citet{chaplin2014}},  and with   
\teff~and $\log g$  as close as possible to the Sun's value (5200~K $<$ \teff $<$  6060~K  and   3.63  $<$ $\log g$ $<$  4.40). This  seismic sample allows a 
direct  comparison  between gyro and seismic--ages for a subset of eight stars.
These  {\it seismic sample}  were observed in short cadence for one month each in  survey mode. 
Stellar properties for these stars have been estimated using two global  asteroseismic  parameters and 
complementary photometric and spectroscopic  observations as described by \blue{\cite{chaplin2014}}.   
The median final quoted uncertainties for the full  \blue{\cite{chaplin2014}}  sample were approximately 0.020 dex in $\log g$   and 150 K in \teff. 

 \vspace{0.3cm}

The 35 selected stars  from the  improved  version of the KIC (\blue{\citealt{brown11}}, \blue{\citealt{huber2013}}) 
present stellar parameters \teff~and $\log g$    in the range   $5700$~K $\leq$  \teff~   $\leq 6120$~K and   $4.3 \leq \log g \leq 4.6$.  
 The median final quoted uncertainties were approximately [+0.070, -0.280]  dex in $\log g$ and 160 K in \teff.
 We derived isochrone ages for these stars. We also included the well--studied solar analogs 16~Cyg~A and B  
 (KIC~12069424 and 12069449; V$\sim$6), with estimated ages of $t=6.8\pm0.4$~Gyr for both stars.
 \blue{ \citep{metcalfe2012,Lund2014}}. 
 So far, for 16~Cyg~A \& B,  there have been no  direct measurements of surface rotation.  
 The age implies that $P_{\mathrm {rot}}$  should be near 30 days  (\blue{\citealt{sku72}}, \blue{\citealt{barnes2010}}) for both components.  Table~\ref{tbl:obs}  summarizes the properties of our  targets.
 The  sample is displayed in Figure~\ref{evoldig_f1}. 
%%%%%%%%%%%%%%%%%%%%%%%

 \vspace{0.3cm}

\section{Evolutionary status}
\label{s_models}
We used  models computed  with the  
 Toulouse-Geneva stellar evolution code \blue{(\citealt{huibonhoa_2008}},
 \blue{\citealt{Richard_1996}};  \blue{\citealt{donascimento2013})}.
 These models used an initial composition with the \blue{\citet{Grevesse_93}} mixture. 
 Rotation-induced mixing and the  transport of angular momentum due to rotationally induced instabilities are computed as described by   \blue{\citet{Talon_1997}},  and these models take into account internal differential rotation. The angular momentum evolution follows the \blue{\citet{Kawaler_1988}} prescription. Our solar model is calibrated to match the observed solar  \teff, luminosity, and rotation at solar age \blue{\citep{Richard_1996}}.  
 Models are shown in Figure~\ref{evoldig_f1} (left). 
\begin{figure}   %%%%%%%%%%%%%%%%%%%%%%%%%%%%%%%%%%%%%%%%% Figure 2  %%%%%%%%%
\centering
\vspace{0.0cm}
\hspace{-0.6cm}
\includegraphics[angle=0,width=8.5cm,height=8.5cm]{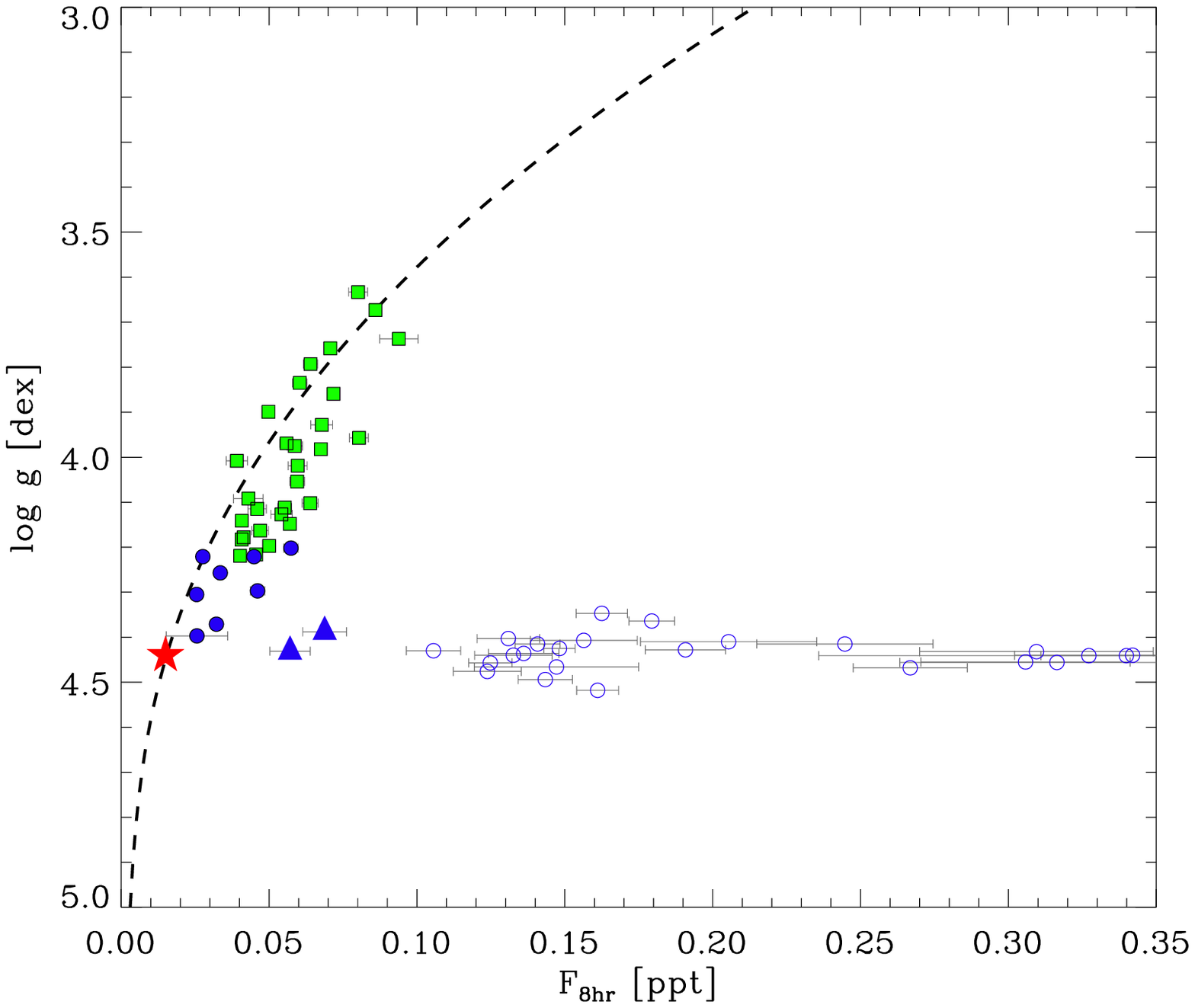}
\vspace{0.0cm}
\caption{Surface gravity from asteroseismology  (filled  squares and circles)  
determined by \blue{\citealt{chaplin2014}} and from the \blue{\cite{huber2013}} (open circles)  
as a function of the 8-hour flicker F$_{8}$, \blue{\citep{bastien13}}. 
Circles represent dwarfs and  squares represent the subgiants. 
The black dashed line corresponds to the relationship derived by \blue{\cite{bastien13}}.
The red symbol at lower left represents the Sun  $\log g$ and  F$_{8}$. Filled triangle represent KIC~2718678, and KIC~12157617.}
\vspace{0.2cm}
\label{figure_flicker}
\end{figure}
We have computed standard (without rotation) and  non-standard models for stellar masses between 0.9 and 1.1 \msun~ 
and for different metallicities consistent with the range of our sample stars. To discriminate between dwarfs and 
subgiants, models are shown with a dashed black  line  that
indicates the  evolutionary point where the subgiant branch starts and which corresponds to  hydrogen exhaustion in 
the stellar core (i.e., turnoff point).   The 30 stars above this line are subgiants.  The remaining 45 stars 
below this line are dwarfs and of primary interest here.  From these, eight are dwarfs with ages known from 
asteroseismology. Finally, we determine the individual masses of our 
sample  of dwarf stars by using the evolutionary tracks for the respective metallicities.
From this, most of the mass determination uncertainty is due to the 
uncertainties on \teff~($\sim$150K) which lead to an error of $\sim$0.05 \msun.
Our determined values  agree with those of  \blue{\cite{huber2013}}.  
For dwarfs, the mass determination is  negligibly affected by the choice of standard 
or non-standard models, and the uncertainties are smaller than intrinsic errors in the $\log~g$ and \teff.
The  mass determination uncertainties  become significant for subgiants. 

\section{Extracting the surface rotation rates}
\label{rot}
The  average surface $P_{\mathrm {rot}}$ is obtained  from light curve modulation analysis.
To extract the $P_{\mathrm {rot}}$,  we analyze  PDC-MAP and  simple aperture photometry 
light curves \blue{\citep{Christiansen2013}} that are corrected for outliers, drifts, and discontinuities and stitched together 
following the procedures described by \blue{\citet{gar11}}. The light curves are then high-pass filtered using a triangular smooth 
function with a cut-off that can change from  40 to 100 days.  We remind readers that the PDC-msMAP \blue{\citep{Christiansen2013}}
corrected data proposed by  \emph{Kepler}  cannot be used here, because 
some quarters are  filtered with a 20 day high-pass filter (\blue{\citealt{gar14}}).
\begin{figure*}   
\centering
\vspace{-0.2cm}
\hspace{0.5cm}
\includegraphics[angle=0,width=10.5cm,height=10.5cm]{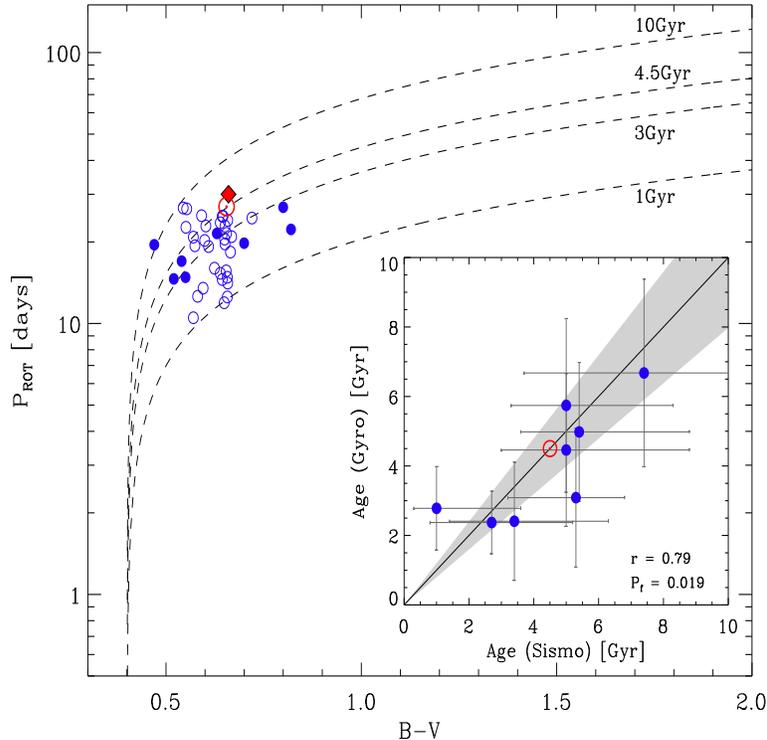} 
\vspace{-0.2cm}
\caption{Color---period diagram for  the seismic sample
for which there are {\it Tycho} photometry (\blue{\citealt{hog2000}}) and solar analogs.
The inside panel shows a comparison between gyro and seismic--ages for  the
{seismic sample} stars  with a shaded region  showing a $\pm~20\%~$ error.}
\label{prot_bv}
\vspace{0.2cm}
\end{figure*}
From a Lomb--Scargle periodogram \blue{\citep{Scargle_1982}}, we searched  for the highest peaks  below 11.57 $\mu$Hz with one daily alias.  To ensure that 
the retrieved peaks are due to stellar signals, we verified that the signal responsible for the   $P_{\mathrm {rot}}$ is present during most 
of the observed  window. Then, we visually examined the results and any period detected 
was considered robust if repeated features were visible in different parts of the light curve.  Finally, we  computed the autocorrelation of 
the  signal (\blue{\citealt{McQuillan2013}}) in order to cross-check the results.   We fit a Gaussian function to the global wavelet 
spectrum \blue{\citep{mathur2010}}, i.e., the projection into the frequency domain of the time frequency analysis and the uncertainty was obtained from the 
full width at half-maximum  of the Gaussian profile. This uncertainty includes the $P_{\mathrm {rot}}$  variation  due to 
differential rotation.  Our entire sample is composed of 75 stars,  
and their derived periods and  errors are presented in Table~1.   

    \vspace{0.2cm}

\section{KIC parameters and flicker measures}
\label{flicker}
In order to better interpret  $P_{\mathrm {rot}}$, it is important to get an idea of 
how reliable  the fundamental  parameters   \blue{\citep{victor2012}} are.
We compared the KIC parameters with those determined by   \blue{\cite{chaplin2014}}  
and with revised  \blue{\cite{huber2013}} values  for all of the stars of the
 {seismic sample} and  solar analogs and  twin candidates,  see Figure~\ref{evoldig_f1} (right).
The \teff~from the KIC is systematically lower when compared to values based on  the
 asteroseismic analysis (Figure~\ref{evoldig_f1}, bottom right panel). \blue{\cite{huber2013}}  incorporated 
 priors results in the distributions of  the parameters at fixed temperature and  used  colors 
 (e.g., $J - K$, $H - K$, $g - i$) with  little sensitivity to [Fe/H] to arrive at more realistic distributions 
 around  0.01 dex in $\log g$. We adopted  the \blue{\cite{huber2013}} values.
 
 \vspace{0.2cm}

 We have measured  the ``flicker'' as defined by \blue{\cite{bastien13}} and 
 validated by \blue{\cite{cranmer2014}}, as a proxy of  the granulation properties that scale with the stellar  $\log g$  \blue{\citep{mathur2011}}. 
 Since we analyzed 1440 days of data, we slightly modify  the method by dividing the time series into  \emph{Kepler} quarters 
 (where Q1 is a month long and the following quarters are each three months)  and we computed the standard deviation of each 
 sub-series smoothed over an eight hour boxcar (the same results were obtained using a six hour function). 
We then corrected the values for the photon noise \blue{\citep{Jen10}}. We applied this procedure to our entire  sample. 
In Figure~\ref{figure_flicker}, we display $\log g$  as a function of the eight hour flicker F$_{8hr}$ which we measured.  
For the filled symbols (dwarfs and subgiants), we can clearly  see the  
 correlation as shown by \blue{\cite{bastien13}}. The red star represents the Sun's  $\log g$, fixed at  $\log g =$ 4.44, and flicker F$_{8hr}$
 \blue{\citep{bastien13}}.  We can distinguish two stars  (filled triangle), namely  KIC~2718678 ($V\sim$~11.49) and 
KIC~12157617 ($V\sim$~11.89), located  close to the position of the Sun.  Based on \blue{\cite{bastien13}}, this plot 
suggests that these two stars are more likely to be very similar to the Sun from their activity level,  as measured by the eight hour 
flicker F$_{8hr}$. A second group of stars has similar values of  $\log g$ and with a wide range of flicker F$_{8hr}$. 
These stars are  represented by open circles belonging to the large-flicker horizontal line shift away from the dashed line. Those 
might be stars with an activity level at the mid-point of their cycle that is larger than the Sun's at solar maximum.
They  might be stars younger than the Sun or stars expected  to have a higher noise (true for faint 
magnitudes,  $K_p > \,13$). We do not expect to find a solar twin when we horizontally move that far from the dashed line.
(F. Bastien 2014, private communication). Our flicker diagram should be encompass  solar dwarfs  at different age-points 
and different  activity levels.  Two stars from  this  sample  are confirmed to be quite similar to 18~Sco  by 
\blue{\citet{Nogami14}}, based on HDS@SUBARU observations.

\section{Rotation--age relationship for solar analogs}
\label{rota_age}

\begin{figure}  
\centering
\vspace{0.0cm}
\hspace{-0.6cm}
\includegraphics[angle=0,width=8.5cm,height=8.5cm]{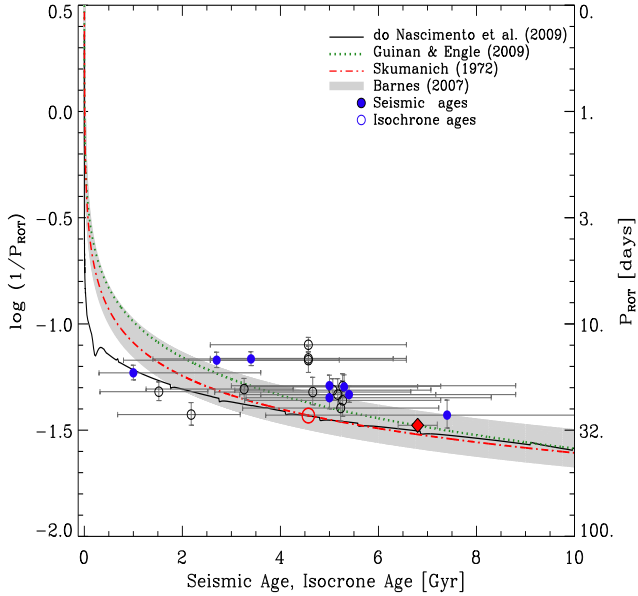} 
\vspace{0.0cm}
\caption{Angular velocity  of  dwarfs with 
1.0 \msun~as a function of  stellar age. Filled circles represent the seismic sample dwarfs. Open circles represent  solar 
twin candidates. The shaded region represents 
the gyro relations  (\blue{\citealt{Barnes07}}) with  $(B-V) = 0.642$ for the Sun, with 
 $0.95~\leq  M/M_{\odot}   \leq~1.10$. Diamonds indicate 16~Cyg~A and B.}
\label{prot_age}
\end{figure}

The rotation rates of these stars permit a deeper understanding of this sample.  In  Figure~\ref{evoldig_f1} we  show the evolution of the
averaged rotation period  $\langle P_{\mathrm {rot}}  \rangle$  for  dwarfs and subgiants.  The $\langle P_{\mathrm {rot}}  \rangle$   
increases  (21$d\rightarrow$26$d\rightarrow$32$d\rightarrow$49$d$) 
as stars evolve up the giant branch.  This figure shows  the broad trend, constrained by sample size, 
of the $P_{\mathrm {rot}}$  of the subgiants. However, our real interest, as far as solar analogs and twins are concerned,
is in understanding the dwarfs. The  $\langle P_{\mathrm {rot}}  \rangle$ for dwarfs   
is slightly lower  than the solar rotation period.  The $P_{\mathrm {rot}}$  that we have measured  from the \emph{Kepler} light curves, together with some solar proxies,  
permit an independent (from classical isochrone or seismic ages) age derivation for the dwarfs using gyrochronology  (\blue{\citealt{Kraft}}; \blue{\citealt{sku72}}; \blue{\citealt{Barnes07}}).    
The basic idea displayed in Figure~\ref{prot_bv} is that the,   $P_{\mathrm {rot}}$ of cool stars are a function of the star's age and 
mass, allowing the age to be determined if the other two variables are known. 
In this work, we bypass the  uncertainties in the color determination 
by working directly with $T_{\rm eff}$, converting it to the  global convective  turnover times $\tau$ using the 
Table~1 in \blue{\citet{Barnes10b}}, and then  calculating the age using  Equation (32) provided in \blue{\citet{barnes2010}}. 
 For $P_{0}$, we used a value of 1.1days, as suggested in \blue{\citet{barnes2010}}.
Remember that, in principle, gyrochronology only provides good results for 
main--sequence stars with masses less than about 1.2 $M_{\odot}$. 
These gyro-ages for dwarfs are listed  in Table~1, along with the uncertainties 
originating in the \teff~ and $P_{\mathrm {rot}}$ uncertainties.
The inset in Figure~\ref{prot_bv}  displays a 
comparison between the  seismic- and gyro--ages on a star-by-star basis. The two ages are in agreement within the errors,
with a correlation coefficient of $r$ = 0.79 with a significance level of $P_{f}$ = 0.019. 
The dwarf ages  range from under 1\,Gyr to ~10\,Gyr, with a median of roughly 3.9\,Gyr, implying that this sample indeed consists 
of stars that are comparable in age to the Sun.   
For 35 dwarf stars from the KIC, we computed  isochrone ages from the model described in  
Section.~\ref{s_models} (Figure~\ref{evoldig_f1}) and gyro-ages (\blue{\citealt{Barnes07}}; \blue{\citealt{mei+11}}).    
Figure~\ref{prot_age} is consistent with the theoretical predictions of \blue{\citet{,saders2013}}. As we expected, 
the isochrone ages  are subject to huge systematic errors, reflected in  the scattered open circles in Figure~\ref{prot_age}, 
coupled with the fact that stellar rotation has a strong dependence on mass. 

\section{Conclusions}
\label{results}

We report rotation periods for 30 subgiants  and 43 dwarfs, where  34 are main-sequence stars with  
$0.90~\leq  M/M_{\odot}   \leq~1.10$. For dwarfs, we  determine individual masses and ages. This resulted in the identification of 
a new sample of at least 22 new solar twin candidates.  Eight of these dwarfs stars (seismic sample) 
have asteroseismic ages  determined by \blue{\cite{chaplin2014}} and allow a direct  comparison  between 
gyro and seismic-ages. The comparison  shows reasonable agreement, mostly within  20$\%$ and
with a correlation coefficient $r$ = 0.79.   $P_{\mathrm {rot}}$ for these samples,  permit a general elucidation of how the  
$\langle P_{\mathrm {rot}}  \rangle$ increase as solar  analogs evolve during the main-sequence and subgiant  
phases.  The sample of dwarfs contains stars with $P_{\mathrm {rot}}$ as slow 
as 27 days, however, the  $P_{\mathrm {rot}}$ histogram distribution for our 43 main-sequence stars 
indicates a peak in 19 days, making  these stars slightly younger than 
the Sun.  This agree with \blue{\cite{Soderblom1985}}, who reported that the Sun is within 1$\sigma$ 
standard deviation of stars of its mass and age. The flicker of these stars is also measured and  provides a 
new additional parameter  in the search for solar twins based on its activity. This can be quite useful in future 
missions such as PLATO (\blue{\citealt{rauer2013}}).
 \acknowledgements
{The authors wish to thank the \emph{Kepler} team. J.D.N. acknowledges  CNPq Universal-B 485880/2012--1 and 
PDE 237139/2012. R.A.G. acknowledges the European Community Seventh Framework Program (FP7/2007-2013) no. 269194 (IRSES/ASK), and a 
CNES/$CoRoT$ grant. S.M. was supported by NASA grant NNX12AE17G.   J.D.N. is grateful to  B. Chaplin and S. Saar for 
discussions and suggestions.  This is a part of \emph{Kepler}  KASC Workpackage PE11.5 led by J.D.N. 
This paper is in memoriam of Giusa Cayrel de Strobel. Astronomer and pioneer in research on solar twins. }

\begin{center}
\newpage
\small
\renewcommand{\arraystretch}{2.0}
\begin{longtable*}{rrrlllrccc}
\caption{Rotation of solar analogs and twin candidates revealed by \emph{KEPLER}}\label{tbl:obs} \\

\hline
\hline
\multicolumn{1}{c}{\sffamily $N^{\b{o}}$}
&\multicolumn{1}{c}{\sffamily $KIC$}
&\multicolumn{1}{c}{\sffamily $V^{(\dagger)}$}
&\multicolumn{1}{c}{\sffamily $T_{eff}$}
&\multicolumn{1}{c}{\sffamily $\log g$}
&\multicolumn{1}{c}{\sffamily $P_{rot}^{(\star)}$}
&\multicolumn{1}{c}{\sffamily $S\_Age, I\_Age$ }
&\multicolumn{1}{c}{\sffamily $Gyro\_Age^{(\star)}$}
&\multicolumn{1}{c}{\sffamily $F_{8hr}^{(\star)}$}
&\multicolumn{1}{c}{\sffamily $Mass^{(\star)}$}\\

\multicolumn{1}{c}{\sffamily $\S$}
&\multicolumn{1}{c}{\sffamily }
&\multicolumn{1}{c}{\sffamily $(mag)$}
&\multicolumn{1}{c}{\sffamily $(K)$}
&\multicolumn{1}{c}{\sffamily $(dex)$}
&\multicolumn{1}{c}{\sffamily $(days)$}
&\multicolumn{1}{c}{\sffamily $(Gyr)$}
&\multicolumn{1}{c}{\sffamily $(Gyr)$}
&\multicolumn{1}{c}{\sffamily $(ppt)$}
&\multicolumn{1}{c}{\sffamily $(M_{\odot})$}\\ [0.1cm]

\hline
\endfirsthead

\hline\hline
\multicolumn{1}{c}{\sffamily $N^{\b{o}}$}
&\multicolumn{1}{c}{\sffamily $KIC$}
&\multicolumn{1}{c}{\sffamily $V^{(\dagger)}$}
&\multicolumn{1}{c}{\sffamily $T_{eff}$}
&\multicolumn{1}{c}{\sffamily $\log g$}
&\multicolumn{1}{c}{\sffamily $P_{rot}^{(\star)}$}
&\multicolumn{1}{c}{\sffamily $S\_Age, I\_Age$ }
&\multicolumn{1}{c}{\sffamily $Gyro\_Age^{(\star)}$}
&\multicolumn{1}{c}{\sffamily $F_{8hr}^{(\star)}$}
&\multicolumn{1}{c}{\sffamily $Mass^{(\star)}$}\\

\multicolumn{1}{c}{\sffamily $\S$}
&\multicolumn{1}{c}{\sffamily }
&\multicolumn{1}{c}{\sffamily $(mag)$}
&\multicolumn{1}{c}{\sffamily $(K)$}
&\multicolumn{1}{c}{\sffamily $(dex)$}
&\multicolumn{1}{c}{\sffamily $(days)$}
&\multicolumn{1}{c}{\sffamily $(Gyr)$}
&\multicolumn{1}{c}{\sffamily $(Gyr)$}
&\multicolumn{1}{c}{\sffamily $(ppt)$}
&\multicolumn{1}{c}{\sffamily $(M_{\odot})$}\\ [0.1cm]

\hline
\endhead
\hline
\multicolumn{2}{c}{\scriptsize\itshape continued}\\
\endfoot
\hline
\endlastfoot
1 & 2718678 & 11.493 & $ 6105 ^{+ 137 } _{- 177 }$ $^{(\sharp)}$ & $ 4.431 ^{+ 0.060 } _{- 0.286 }$ $^{(\sharp)}$ & $ 24.5 ^{+ 2.50 } _{- 2.50 }$  & $ 1.9 ^{+ 1.0 } _{- 1.5 }$ $^{(\star)}$ & $ 7.91 ^{+ 4.0 } _{- 4.0 }$ & $ 0.057 ^{+ 0.007 } _{- 0.007 }$ & $ 1.03 ^{+ 0.07 } _{- 0.08 }$  \\
2 & 3118654 & 13.235 & $ 5775 ^{+ 154 } _{- 157 }$ $^{(\sharp)}$ & $ 4.407 ^{+ 0.090 } _{- 0.260 }$ $^{(\sharp)}$ & $ 15.6 ^{+ 1.51 } _{- 1.51 }$  & $ 6.2 ^{+ 2.0 } _{- 2.0 }$ $^{(\star)}$ & $ 1.87 ^{+ 0.5 } _{- 0.5 }$ & $ 0.156 ^{+ 0.018 } _{- 0.018 }$ & $ 0.99 ^{+ 0.08 } _{- 0.07 }$  \\
3 & 4473226 & 15.707 & $ 5776 ^{+ 172 } _{- 169 }$ $^{(\sharp)}$ & $ 4.456 ^{+ 0.071 } _{- 0.269 }$ $^{(\sharp)}$ & $ 15.3 ^{+ 1.27 } _{- 1.27 }$  & $ 3.5 ^{+ 2.0 } _{- 2.0 }$ $^{(\star)}$ & $ 1.80 ^{+ 0.4 } _{- 0.4 }$ & $ 0.493 ^{+ 0.051 } _{- 26.910 }$ & $ 0.90 ^{+ 0.10 } _{- 0.10 }$  \\
4 & 5084157 & 11.649 & $ 6054 ^{+ 137 } _{- 137 }$ $^{(\phi)}$ & $ 4.202 ^{+ 0.018 } _{- 0.019 }$ $^{(\phi)}$ & $ 22.3 ^{+ 2.85 } _{- 2.85 }$ $^{(\chi)}$ & $ 5.0 ^{+ 3.3 } _{- 1.7 }$ $^{(\phi)}$ & $ 5.74 ^{+ 2.5 } _{- 2.5 }$ & $ 0.057 ^{+ 0.002 } _{- 0.002 }$ & $ 1.12 ^{+ 0.10 } _{- 0.08 }$  \\
5 & 5184732 & 8.165 & $ 5818 ^{+ 190 } _{- 190 }$ $^{(\phi)}$ & $ 4.257 ^{+ 0.012 } _{- 0.012 }$ $^{(\phi)}$ & $ 19.8 ^{+ 2.43 } _{- 2.43 }$  & $ 5.3 ^{+ 1.5 } _{- 2.1 }$ $^{(\phi)}$ & $ 3.09 ^{+ 1.1 } _{- 1.1 }$ & $ 0.033 ^{+ 0.001 } _{- 0.001 }$ & $ 0.99 ^{+ 0.10 } _{- 0.08 }$  \\
[0.1cm]
\hline
\end{longtable*}

\begin{minipage}{17.5cm}
\tablecomments{\small{Table~\ref{tbl:obs} is published in its entirety in the electronic edition of ApJL. A portion is shown here for guidance regarding its form and content.$^{(\star)}$ This paper; $^{(\dagger)}$ \emph{Kepler} Input Catalogue; $^{(\phi)}$ \blue{\cite{chaplin2014}}; $^{(\sharp)}$ \blue{\cite{huber2013}}; $^{(\flat)}$  \blue{ \cite{metcalfe2012}}.; $^{(\chi)}$ half period detected; $^{(\psi)}$ twice  period detected; $^{(\mu)}$ low modulation; $^{(\rho)}$ bad correction}}
\end{minipage}

%\tablenotetext{a}{Table~\ref{tbl:obs}   is published in its entirety
 % in the electronic edition of ApJ. A portion is shown here for
 % guidance regarding its form and content.}

\end{center}


\begin{thebibliography}{36}

\expandafter\ifx\csname natexlab\endcsname\relax\def\natexlab#1{#1}\fi


\bibitem[{{Baglin~}\it{et~al.}(2006){Baglin}, {Auvergne}, {Barge}, {Deleuil},
  {Catala}, {Michel}, {Weiss}, \& {The COROT Team}}]{baglin06}
{Baglin}, A.,  P., {et~al.} 2006, in ESA Pub., Vol. 1306, 33

\bibitem[{Barnes}(2007)]{Barnes07}
{Barnes}, S. A. 2007, \apj, 669, 1167

\bibitem[{{Barnes}(2010){Barnes}}]{barnes2010}
{Barnes}, S. A. 2010, \apj, 722, 222

\bibitem[{Barnes} \& {Kim}(2010b)]{Barnes10b}
{Barnes}, S. A., \& {Kim}, Y.-C. 2010b, \apj, 721, 675

\bibitem[{{Basri~}\it{et~al.}(2011){Basri}, {Walkowicz}, {Batalha}, {Gilliland},
  {Jenkins}, {Borucki}, {Koch}, {Caldwell}, {Dupree}, {Latham}, {Marcy},
  {Meibom}, \& {Brown}}]{bas+11}
{Basri}, G., {Walkowicz}, L.~M., {Batalha}, N., {et~al.} 2011, \aj, 141, 20

\bibitem[{Bastien~}\it{et~al.}(2013)]{bastien13}
{Bastien}, F.~A.,  {Stassun}, K. G., {et~al.} 2013, Nature, 500,  427


\bibitem[{Bazot~}\it{et~al.}(2011)]{bazot} 
{Bazot}, M., {Ireland}, M.~J., {Huber}, D., {et~al.}\ 2011, \aap, 526, L4

\bibitem[{Borucki~}\it{et~al.}(2010)]{Borucki}
{Borucki}, W.~J., {Koch}, D., {Basri}, G., {et~al.}\ 2010, Science, 327, 977


\bibitem[{Brown~}\it{et~al.}(2011)]{brown11}
{Brown}, T.~M., {Latham}, D.~W.,  {et~al.}\ 2011, \aj, 142, 112


\bibitem[{{Cayrel de Strobel}(1996)}]{cayrel1996}
{Cayrel de Strobel}, G. 1996, A\&A Rev., 7, 243

\bibitem[{Chaplin~}\it{et~al.}(2014)]{chaplin2014}
{Chaplin}, W. J., {Basu}, S., {Huber}, D., {et~al.}  2014, \apjs, 210, 1


\bibitem[{Christiansen~}\it{et~al.}(2013)]{Christiansen2013}
{Christiansen}, J. L., {Van~Cleve}, J. E., {Jenkins}, J. M., {et~al.}~2013,   \emph{Kepler}  Data Characteristics Handbook, KSCI-19040-004


\bibitem[{Cranmer~}\it{et~al.}(2014)]{cranmer2014} 
{Cranmer}, S.~R., {Bastien}, F.~A., {et~al.}\ 2014, \apj, 781, 124


\bibitem[{{do~Nascimento~}\it{et~al.}(2009)}]{Nascimento_2009}
{do~Nascimento}, J. -D., Jr., {Castro}, M.,  {et~al.}  2009, \aap, 501, 687

\bibitem[{{do~Nascimento}\it{~et~al.}(2013)}]{donascimento2013}	
{do~Nascimento}, J. -D., Jr.,  {Takeda}, Y.,   {et~al.}  2013, \apjl, 771, 31


\bibitem[{Dorren~}\&{~Guinan}(1994)]{dorren_guinan1994}
{Dorren}, J. D., \& {Guinan}, E. F. 1994, \apj, 428, 805


\bibitem[{Garc\'ia~}\it{et~al.}(2011)]{gar11} 
{Garc\'ia}, R. A., {Hekker}, S., {Stello}, D., {et~al.} 2011,  MNRAS, 414, 6

\bibitem[{Garc\'ia~}\it{et~al.}(2014)]{gar14}
{Garc\'ia}, R. A., {Ceillier}, T., {et~al.}  2014, arXiv1403.7155



%\bibitem[{Gilliland~}\it{et~al.}(2011)]{gil11} 
%{Gilliland}, R.~L., {Chaplin}, W.~J., {et~al.}\ 2011, \apjs, 197, 6


\bibitem[{{Grevesse~}{\&~}{Noels}(1993)}]{Grevesse_93}
{Grevesse}, N., \& {Noels}, A. 1993, Origin and Evolution of the Elements,
eds.  N. Prantzos, E. Vangioni--Flam, and M. Cass\'e, Cambridge Univ. Press, p.~15


\bibitem[{{Guinan~}{\&~}{Engle}(2009)}]{Guinan2009}
{Guinan}, E. F., \& {Engle}, S. G. 2009, IAUS, 258, 395



\bibitem[{Gustafsson~}(2008)]{gusta2008} 
{Gustafsson}, B.\ 2008, Physica Scripta Volume T, 130, 014036


\bibitem[{{Hardorp}{}(1978){Hardorp}}]{hardorp1978}
 {Hardorp}, J. 1978, A\&A, 63, 383

\bibitem[{H{\o}g~}\it{et~al.}(2000)]{hog2000} 
{H{\o}g}, E.,  {Fabricius}, C., {Makarov}, V. V.,  {et~al.} 2000,  A\&A, 355, 27


\bibitem[{Huber~}\it{et~al.}(2014)]{huber2013}
{Huber}, D.,  {Silva Aguirre}, V.,  {et~al.} 2014, \apjs, 211, 2


\bibitem[{{Hui-Bon-Hoa}(2008)}]{huibonhoa_2008}
{Hui-Bon-Hoa}, A. 2008, Ap\&SS, 316, 55


\bibitem[{Jenkins~}\it{et~al.}(2011)]{Jen10} 
{Jenkins}, J.~M., {Caldwell}, D.~A., {et~al.}\ 2010, \apjl, 197, 6


\bibitem[{{Kawaler~}(1988)}]{Kawaler_1988}
{Kawaler}, S. D. 1988, \apj, 333, 236

\bibitem[\protect\citeauthoryear{Kraft}{1967}]{Kraft}
{Kraft}, R. P. 1967, ApJ, 150, 551


\bibitem[{{Lund~}\it{et~al.}(2014){Lund}, {Kjeldsen}, {Christensen-Dalsgaard}, 
{Handberg}, \& {Silva Aguirre}}]{Lund2014} {Lund}, M. N.,  {Kjeldsen}, H., 
 {et~al.}  2014, \apj, 782, 2

\bibitem[{Mathur~}\it{et~al.}(2010)]{mathur2010}
{Mathur}, S., {Garc\'ia}, R. A., {R\'egulo}, C., {et~al.} 2010, A\&A, 511, 46

\bibitem[{Mathur~}\it{et~al.}(2011)]{mathur2011}
{Mathur}, S., {Hekker}, S., {et~al.} 2011, \apj, 741, 119


\bibitem[{{McQuillan~}\it{et~al.}(2013){McQuillan}, {Mazeh}, \&
  {Aigrain}}]{McQuillan2013} {McQuillan}, A.,  {Mazeh}, T., {et~al.} S. 2013, \apjl, 775, 11



\bibitem[{{Meibom~}\it{et~al.}(2011){Meibom}, {Barnes}, {Latham}, {Batalha},
  {Borucki}, {Koch}, {Basri}, {Walkowicz}, {Janes}, {Jenkins}, {Van Cleve},
  {Haas}, {Bryson}, {Dupree}, {Furesz}, {Szentgyorgyi}, {Buchhave}, {Clarke},
  {Twicken}, \& {Quintana}}]{mei+11}
{Meibom}, S., {Barnes}, S.~A., {et~al.} 2011, \apjl, 733, L9


\bibitem[{{Mel\'endez \& Ram\'irez}{}(2007){Mel\'endez}, {Ram\'irez}}]{melendez2007}
{Mel\'endez}, J., \& {Ram\'irez}, I. 2007, \apj, 669, L89


\bibitem[{{Metcalfe~}\it{et~al.}(2012){Metcalfe}, {Chaplin}, \& {Appourchaux}}]{metcalfe2012}
{Metcalfe}, T. S., {Chaplin}, W.J., {et~al.}  2012, \apj, 748, 10L


\bibitem[{Monroe~}\it{et~al.}(2013)]{Monroe13} 
{Monroe}, T. R., {Mel\'endez}, J., {et~al.} 2013, \apjl, 774L, 32


\bibitem[{Nogami~}\it{et~al.}(2014)]{Nogami14}
{Nogami}, D., {Notsu}, Y.,  {et~al.}  2014, 2014arXiv1402.3772N


\bibitem[{Noyes~}\it{et~al.}(1984)]{Noyes84}
{Noyes}, R. W., {Hartmann}, L. W.,  {et~al.} 1984, \apj, 279, 763


\bibitem[{{Porto de Mello \& da Silva}{}(1997){Porto de Melo}, {da Silva}}]{portomello1997}
{Porto de Mello}, G. F., \& {da Silva}, L. 1997, \apj, 482, L89

\bibitem[{Ram\'irez~}\it{et~al.}(2012)]{ivan2012} 
{Ram{\'{\i}}rez}, I., {Michel}, R., {Sefako}, R., {et~al.}\ 2012, \apj, 752, 5


\bibitem[{Rauer~}\it{et~al.}(2013)]{rauer2013} 
{Rauer}, H., {Catala}, C., {Aerts}, C., {et~al.}\ 2013, arXiv1310.0696


\bibitem[{{Richard~}\it{et~al.}(1996)}]{Richard_1996}
{Richard}, O., {Vauclair}, S.,  {et~al.} 1996, \aap, 312, 1000

\bibitem[{{Scargle}(1982)}]{Scargle_1982}
{Scargle}, J. D. 1982, \apj, 263, 835


\bibitem[{Silva Aguirre~}\it{et~al.}(2012)]{victor2012}
{Silva Aguirre}, V., Casagrande, L., {et~al.}  2012, \apj, 757, 99 


\bibitem[{Skumanich~}(1972)]{sku72} 
{Skumanich}, A.\ 1972, \apj, 171, 565


\bibitem[{{Soderblom~}(1985){Soderblom}}]{Soderblom1985}
{Soderblom}, D. R. 1985, \aj, 90, 2103


\bibitem[{Soderblom}(2010)]{Soderblom10}
{Soderblom}, D. R. 2010, ARA\&A, 48, 581 


%\bibitem[{Soubiran} \& {Triaud}(2004)]{soubiran}
%{Soubiran}, C., \& {Triaud}, A.\ 2004, \aap, 418, 1089

\bibitem[{{Talon~}{\&~}{Zahn}(1997)}]{Talon_1997}
{Talon}, S., {Zahn}, J.-P. 1997, \aap, 317, 749

\bibitem[{{van Saders~}{\&~}{Pinsonneault}(2013)}]{saders2013}
{van Saders}, J., \& {Pinsonneault}, M. 2013, \apj, 776, 67

\bibitem[{Wilson~}(1963)]{wilson1963} 
{Wilson}, O.~C. 1963, \apj, 138, 832
\end{thebibliography}
\end{document}